# Efficient Topology Optimized Couplers for On-Chip Single-photon Sources

Omer Yesilyurt[1,2,#], Zhaxylyk A. Kudyshev[1,2,#], Alexandra Boltasseva[1,2],
Vladimir M. Shalaev[1,2], and Alexander V. Kildishev[1]

[1]School of Electrical and Computer Engineering, Birck Nanotechnology Center and Purdue Quantum Science and Engineering Institute, Purdue University, West Lafayette, IN 47907, USA

[2]The Quantum Science Center (QSC), a National Quantum Information Science Research Center of the U.S. Department of Energy (DOE), Oak Ridge, TN, 37931, USA, Oak Ridge, TN 37931, USA

[#]Authors with equal contributions

**ABSTRACT:** Room temperature single-photon sources (SPSs) are critical for the emerging practical quantum applications such as on-chip photonic circuity for quantum communications systems and integrated quantum sensors. However, direct integration of an SPS into on-chip photonic systems remains challenging due to low coupling efficiencies between the SPS and the photonic circuitry that often involve size mismatch and dissimilar materials. Here, we develop an adjoint topology optimization scheme to design high-efficiency couplers between a photonic waveguide and SPS in hexagonal boron nitride (hBN). The algorithm accounts for fabrication constraints and the SPS location uncertainty. First, a library of designs for the different positions of the hBN flake containing an SPS with respect to a $Si_3N_4$ waveguide is generated, demonstrating an average coupling efficiency of 78%. Then, the designs are inspected with dimensionality reduction technique to investigate the relationship between the device geometry (topology) and performance. The fundamental, physics-based intuition gained from this approach could enable the design of high-performance quantum devices.

INTRODUCTION

Photonics is a promising route to practical quantum information processing applications because it offers high speed, tolerance to decoherence in transparent environments, and high internal degrees of freedom, which can be used for quantum state encoding.[1] Leveraging the recent advances in classical photonic integrated circuits, various elements of quantum photonic integrated circuits (QPICs)[2–5] have already been demonstrated at a proof-of-concept level. The realization of scalable QPICs is a critical milestone on the way to practical quantum photonic computing,[6] quantum information processing applications,[1] quantum communication,[7,8] quantum sensing,[9] quantum key distribution[10] and machine learning.[11] The operation of the QPICs consists of three main parts: (i) generation of pre-defined quantum states, (ii) transformation of the quantum states via linear and nonlinear on-chip photonic elements, and (iii) quantum state readout.[12] Although different components of the QPICs have been demonstrated, there is not yet a clear pathway for practical realization of scalable QPICs, which requires the development of highly efficient, on-chip components for each of the parts of QPIC. An essential aspect of QPIC designs is compatibility with large-scale fabrication techniques, such as e-beam, photolithography, and etching, in terms of tolerance against the imperfections introduced during the fabrication process. In order to address the highly constrained optimization problem of QPIC design, advanced optimization frameworks conventionally used to solve a broad range of design problems in classical photonics could be employed. There are two main paradigms for the integrated photonic component development: (i) methods based on prior knowledge or scientific intuition and (ii) algorithm-driven inverse design optimization frameworks.[13] The designs obtained with the first approach demonstrate simplified geometrical shapes and intuitive underlying physics, along with substantially limited performance due to reduced internal degrees of freedom. The second approach finds the solution to the inverse problems by determining the material distribution, which ensures the desired optimal performance of the photonic device. The inverse algorithms lead to devices with non-trivial shapes and topologies, enabling high performance and compatibility with the pre-defined fabrication precisions. The inverse design optimization frameworks have been applied to traditional optimization problems in photonics, such as optimizing silicon photonic circuitry elements,[14–16] photonic crystal-based structures,[17] and metasurfaces.[18–20] Recently, topology optimization[21,22] (TO) was applied to various inverse design problems for quantum photonic applications. Namely, TO has been applied to realize efficient and coherent light-matter interfaces via integration of nitrogen-vacancy (NV) and tin-vacancy centers based SPSs within the diamond material platform.[23] TO has been used for controlling the emission of near-surface NVs in diamond with optimized silicon[24] and gallium phosphide meta-structures.[21] Within this work, we optimize the single-photon emission coupling efficiency of a point defect in a hexagonal boron nitride (hBN) flake into a pre-defined fundamental mode of the silicon nitride ($Si_3N_4$) waveguide. Specifically, we apply an adjoint TO framework to optimize the SPS coupler to enable (i) high-efficiency emission coupling into the $Si_3N_4$ waveguide and (ii) robustness of the design against perturbations introduced by fabrication and uncertainty of the emitter position inside the hBN flake. Finally, we implement dimensionality reduction analysis on the TO designs, intending to gain an insight into the physics behind the complex geometrical structures that leads to high-efficiency coupling. We demonstrate that such a rigorous analysis of the topology-optimized sets generates the initial material distribution, equivalent to an "educated guess" obtained with physics-driven human intuition. The proposed approach, which could take one from the "human-sketched" concept to "computer-perfected" design of high-efficiency SPS couplers substantially faster, can be extended to other inverse design problems in photonics.

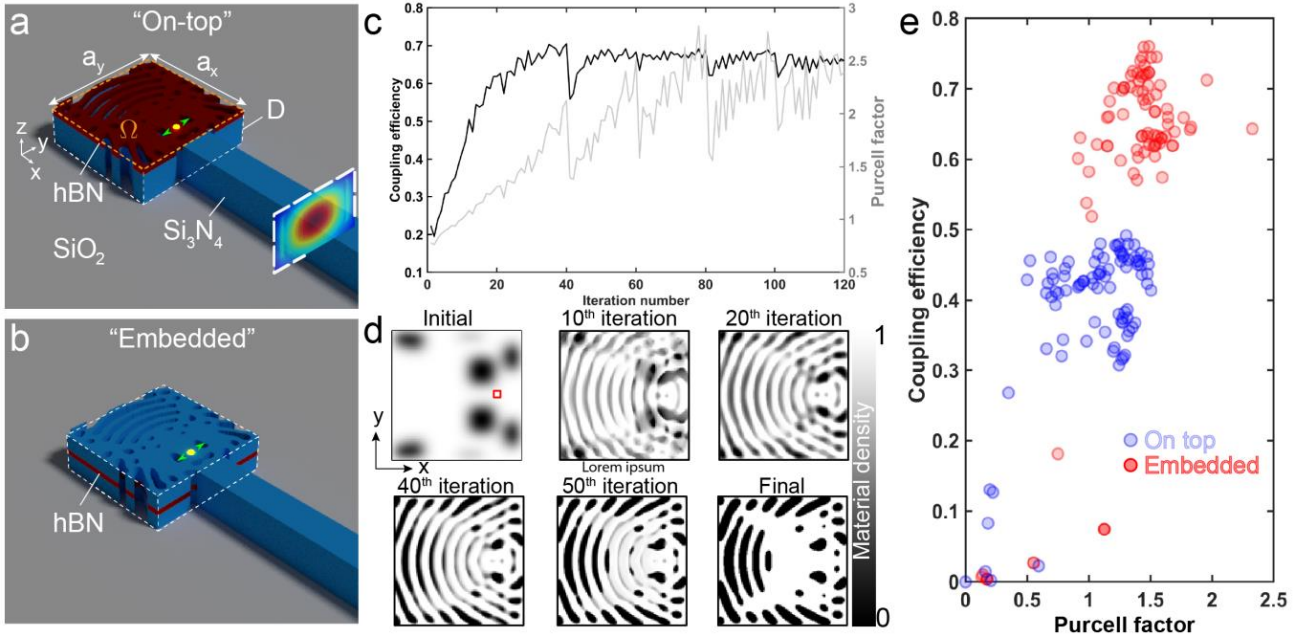

**Figure 1.** Topology optimization of the $Si_3N_4$-hBN hybrid coupler cavity. (a)-(b) Schematics of the optimization domain $D$ for "on top" (a) and "embedded" (b) configurations of $Si_3N_4$-hBN hybrid cavity on $SiO_2$ substrate. The lateral cross-section $\Omega$ of D with an area of $a_x \times a_y$ ($a_x = 2$ μm, $a_y = 2$ μm) is highlighted by the dashed orange box in (a). The total thickness of the optimized cavity is 200 nm with a 50-nm-thick hBN and a 150-nm-thick $Si_3N_4$ layers. SPS is assumed to be a point dipole source with its vector electric dipole moment aligned along $y$-axis and emitting at a wavelength of 638 nm. Position and orientation of the dipole moment are shown with the yellow marker. (c) Evolution of the coupling efficiency (black) and the Purcell factor (gray) during a TO run. (d) Corresponding evolution of the material density distribution inside the optimization area. Position of the SPS is shown with the red marker. The waveguide is assumed to be on the right side of the optimization area. (e) Distribution of the efficiencies vs the Purcell factor for each of the designs for "on top" (blue) and "embedded" (red) configurations.

EFFICIENT CONVERSION OF SPS EMISSION VIA TOPOLOGY OPTIMIZED COUPLER

One of the possible ways of coupling SPSs into an on-chip infrastructure is to use evanescent coupling[25]. Recently, it has been demonstrated that placing an hBN flake with a color center on top of an aluminum nitride waveguide provides up to 1.2% extraction efficiency through the grating out-coupler.[26] Such a limited performance results from low hBN-waveguide coupling efficiency (15.5%, numerical analysis) and grating out-coupler extraction efficiency (9.8%, measurements). The deterministic integration of the SPSs in hBN into $Si_3N_4$ waveguides has also been demonstrated.[27] It has been theoretically predicted that such an approach could yield ~20% coupling efficiency. Since the performance of such a hybrid system could be substantially improved by utilizing an inverse design optimization framework, we apply adjoint TO to design a high-efficiency coupler for an hBN-based SPS and a $Si_3N_4$ waveguide. Specifically, we develop an inverse-designed $Si_3N_4$-hBN hybrid coupler that provides high-efficiency conversion of SPS emission into the fundamental TM mode of 500×200 $nm^2$ $Si_3N_4$ waveguide placed on an $SiO_2$ substrate.

In the developed quantum coupler design (Figure 1a), the optimization domain is a 2×2 $μm^2$ cross-section area made of $Si_3N_4$ and hBN layers with identical topology. The SPS is assumed to be a point dipole source embedded inside the hBN layer, with its electric dipole moment aligned along the $y$-axis and emitting at a wavelength of 638 nm (the zero-phonon line of color centers in hBN). Two different configurations are considered: "on-top" – 50 nm thick hBN is placed on top of the 150 nm thick $Si_3N_4$ layer (Fig. 1a), and "embedded" – the hBN slab is embedded between two 75 nm thick $Si_3N_4$ layers (Fig. 1b). One of the lateral sides of the optimization domain is connected with the Si₃N₄ waveguide, and the propagation direction is assumed to be along the *x*-axis.

The figure of merit (FoM) of the coupler is defined as a ratio between the fraction of the power emitted by the dipole source into the fundamental TEM mode and the total power emitted by the SPS, which is also denoted here as coupling efficiency. Here, we used an adjoint variation of the TO algorithm with sensitivity analysis, which enables the robustness control of optimized designs against the lateral perturbations of SPS position introduced due to limited fabrication precision.[28,29]

The adjoint TO is realized via two full-wave simulations per optimization iteration, forward and adjoint, and employs a commercial finite-difference time-domain (FDTD) solver (Lumerical FDTD). The forward simulation is performed with dipole source excitation within the coupler domain $D$ connected to the waveguide. In contrast, the adjoint simulation is conducted via the TM mode excitation in the backward direction. The dipole source (forward simulation), and spatial distribution of the TM mode excitation used for the adjoint run, are shown in Fig. 1a. $E$-field distributions within $D$ retrieved from forward and adjoint runs are then used to calculate the FoM gradient distribution in $D$. The FoM gradient values are used for updating the material density distributions in xy-cross-section $\Omega$ for the next iteration. More details on the adjoint TO framework can be found in Supporting Information, Section S3. The choice of the initial material distribution plays an essential role in the adjoint TO framework. Initially, the material distribution in the optimization domain is set to be a random, smooth distribution with the following permittivity function,

$$\varepsilon(x,y) = \varepsilon_{air}\left[1 - \rho(x,y)\right] + \varepsilon_{mat}\rho(x,y), \quad \rho \in [0,1] \tag{1}$$

where $\rho(x,y)$ is the material density distribution that varies from 0 (air) to 1 (material), $\varepsilon_{air}$, $\varepsilon_{mat}$ are the dielectric constants of air and material domain (Si₃N₄ and hBN), respectively. Material distribution along the *z*-axis is a uniform translation of the cross-section map, and corresponding permittivity values are used within the Si₃N₄ and hBN layers.

Figure 1c shows the convergence dynamics for one of the TO runs, depicting the evolution of coupling efficiency during the optimization run. The coupling efficiency is gradually increasing and saturates around a steady-state local solution determined by the initial material distribution and parameters of the TO, such as parameters of the binary push and filtering procedures[30] (see Supporting Information, Section S3). The spikes in the FOM convergence plot are caused by the filtering algorithm suppressing sub-precision features.

The optimization starts with a randomly selected initial material distribution and gradually converges to a binary material pattern at the very end of the optimization cycle (Fig. 1d). We intentionally constrain material distribution around the dipole source position to a material density of the SiN/hBN to eliminate voids around the dipole (see highlighted red box in Fig. 1d).

Along with the coupling efficiency of the SPS into the waveguide, the Purcell factor, which defines the rate of the SPS's spontaneous emission, plays an important role. Previously, it has been demonstrated that the Purcell factor can be substantially increased via placing the SPS in a suitable photonic/plasmonic environment with an increased electromagnetic local density of states.[31,32] Although, in this work, TO is used for optimization of the coupling efficiency; the Purcell factor optimization can also be explicitly incorporated into the TO framework. During the TO process, along with the increase of the coupling efficiency, a ~2.5× increase of the Purcell factor of the structure is achieved (Fig. 1c, grey curve). Such an increase of the Purcell factor is mainly driven by the rise in the quality factor of the Si₃N₄-hBN coupler. Such a moderate increase is the result of the constrained uniform material distribution around the dipole source. By unconstrained optimization of the coupler design, it is possible to substantially increase the Purcell factor (~15) while maintaining high coupling efficiency. However, such an approach might lead to impractical coupler designs containing air voids around the dipole position and need additional postselection of the optimized structures. Based on the developed TO framework, we conducted 85 optimization runs for each of the "on-top" and "embedded" geometrical configurations (Fig. 1e). The "embedded" configuration provides more efficient coupler designs in terms of both coupling efficiencies (76%) and Purcell factor (2.5), while the best design in the "on top" set yields only 49% coupling efficiency and 1.5× Purcell factor (Fig. 1e).

OFF-CENTER CONFIGURATION AND ROBUSTNESS.

In the previous section, TO is realized for a fixed position of the SPS dipoles. However, in reality, it is challenging to realize. Moreover, due to the limited precision of the localization of point defects in hBN flakes, the coupler designs should be robust against the uncertainty of the SPS position within the Si₃N₄-hBN coupler.

In this section, we showcase the off-center configuration of the coupler and incorporate the robustness procedure into the TO algorithm against perturbations due to imperfect SPS positioning. For the off-center configuration, the position of the dipole is randomly picked within the optimization domain. Specifically, we consider four different positions of the dipole (red markers, Fig. 2a). Corresponding field distributions and resulting coupler designs are also shown in Fig. 2a. Here, we consider "embedded" designs configurations for all off-center cases. The field distributions prove that optimized cavities indeed enable the higher efficiency conversion of the emission into the waveguide mode. From the coupler designs, it can be seen that the TO leads to the formation of the "Bragg grating" type material distribution around the SPS, which leads to more efficient redirection of the field towards the waveguide. The off-center position of the SPS leads to a reduction of the coupling efficiency (< 55%) while providing relatively high values of the Purcell factor (Fig. 2a). Such reduction of the coupling efficiency occurs due to a relatively small refractive index of $Si_3N_4$. One possible way of overcoming this issue is to increase the dimensions of the $Si_3N_4$-hBN coupler, which would allow to redirection of the SPS emission into the waveguide adiabatically. On the other hand, the same simulation examples show that off-center cases provide a relatively large Purcell factor. Such enhancement of the Purcell factor depends on the material distribution around the SPS source. Within the considered off-center cases, the Purcell factor increases due to the formation of air voids around the SPS, i.e., due to the formation of a dielectric cavity with a high quality factor. Yet another vital aspect is the robustness of $Si_3N_4$-hBN couplers against the uncertainty in the point defect position within the hBN layer. Such uncertainty occurs due to the limited precision of the SPSs' positioning within the hBN flake. To include such a constraint, we implemented a maximin decision rule into the TO framework (see, Supporting Information, Section S3). For each TO iteration, the dipole source was placed into randomly chosen positions within a pre-defined area. Corresponding FOMs are assessed by conducting forward simulation runs. Configuration with the lowest FOM value is used to determine material gradient distribution and updating the material distribution for the next iteration. Such a maximin technique allows for convergence to a coupler design which is robust against the uncertainty of the positioning of the SPS within the pre-defined area (Fig. 2b). Along with the evolution of the FOM, the Purcell factor is evolving as well along the TO run (Fig. 2b) that eventually yields a robust cavity with 66% coupling efficiency and 2.4 Purcell factor for the case of 50×50 nm$^2$ optimization area.

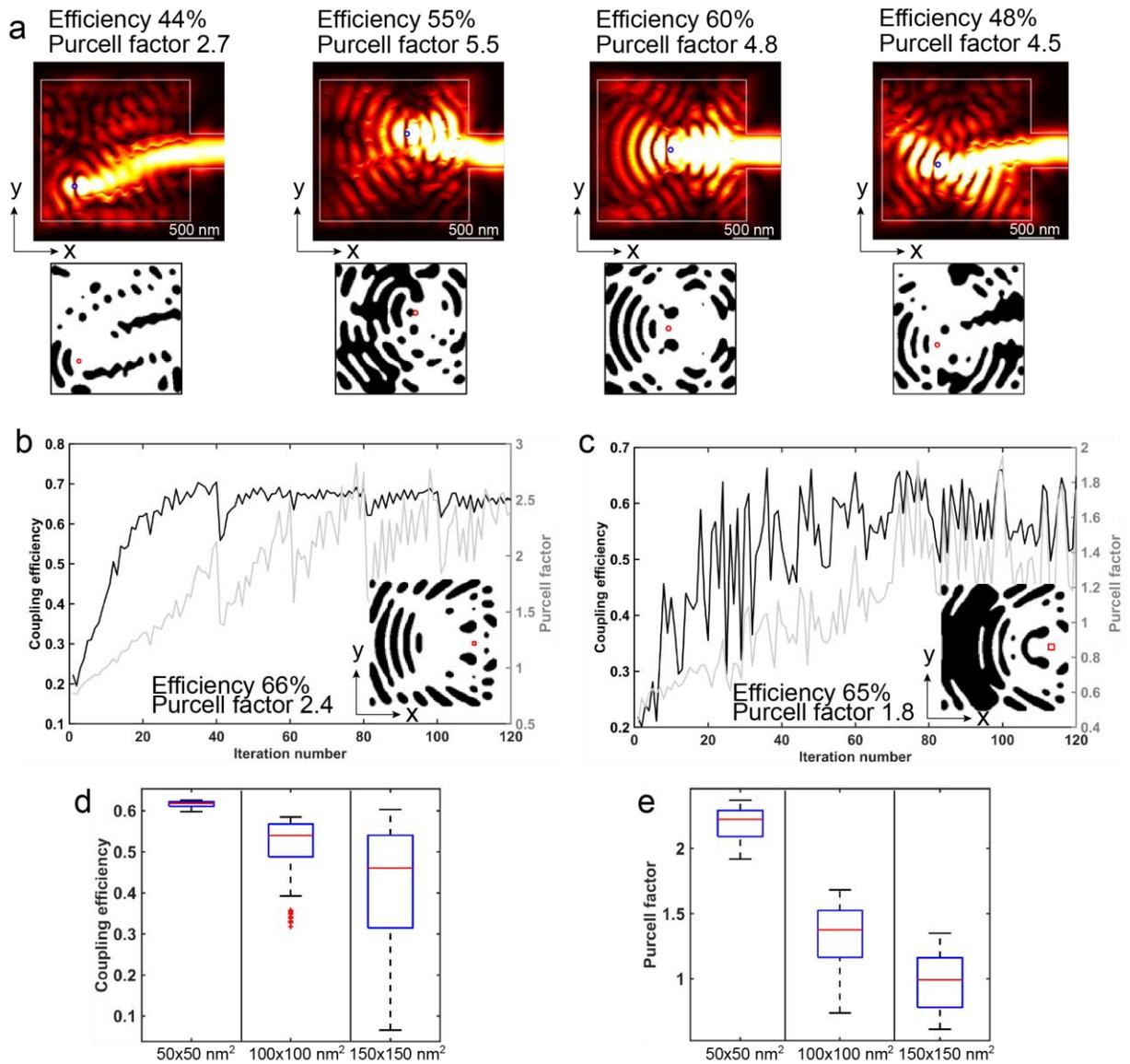

**Figure 2.** Robustness of the $Si_3N_4$-hBN coupler against uncertainty of the SPE position. (a) Field distribution (top) and final designs (bottom) of the coupler for off-center position. Position of the dipole source is highlighted by red marker. (b)-(c) Convergence plots of the TO with maximin decision rule for the robustness against uncertainty of the point defect position within $50×50$ $nm^2$ (c) and $100×100$ $nm^2$ (d) area. Black curve: FOM, grey curve: Purcell factor. Insets show final design of the $Si_3N_4$-hBN coupler. (d)-(e) Coupling efficiency(d) and the Purcell factor(e) distributions compared for the designs optimized with maximin decision rule for the robustness against perturbations over the areas of $50×50$ $nm^2$, $100×100$ $nm^2$, and $150×150$ $nm^2$. The boxed plot shows the median (red line), interquartile range (box), and outliers (red markers).

We also studied the convergence of FOM and the Purcell factor for the case of SPS position perturbation within 100×100 nm$^2$ area (Fig. 2c). In this case, the TO ensures 65% coupling efficiency and 1.8 Purcell factor. One should note that the larger perturbation areas of the SPS position perturbation lead to unstable convergence of the TO. Both cases were realized via probing three different SPS positions at each iteration. Hence, each TO iteration is realized via 4 full-wave runs in total (3 forward and 1 adjoint FDTD runs). To test the performance of the final devices, we optimized three structures with a maximin decision rule applied to 50×50 nm$^2$, 100×100 nm$^2$, and 150×150 nm$^2$ areas. Position of the source dipole is swept in 10 nm steps along both lateral directions in the pre-selected areas of the optimized devices, and corresponding coupling efficiencies, as well as Purcell factors, are retrieved. Statistics of coupling efficiency and Purcell factors are shown in Fig. 2d-e, respectively. The TO with maximin decision rule applied for 50×50 nm$^2$ area ensures highly robust performance of the device with the median of FOM at 61.8% and interquartile range (25$^{th}$ to 75$^{th}$ percentile) between 61% and 62.2%. For comparison, the 100×100 nm$^2$ robustness case has the median at 54% and interquartile range between 48.8% and 56.8%, while the 150×150 nm$^2$ case has the median at 46% and interquartile range between 31.5% and 54%. The Purcell factor distributions show the same behavior. While the 50×50 nm$^2$ area case shows a narrow distribution of the Purcell factor (median at 2.22 and interquartile range between 2.09 and 2.29), the 100×100 nm$^2$ and 150×150 nm$^2$ cases exhibit distributions with medians at 1.37 and 1 respectively. From this analysis, we conclude that the 50×50 nm$^2$ area case shows the most robust performance of the coupler with a deviation of the coupling efficiency within 1.2%. The 100×100 nm$^2$ ensures quite good stability of coupling efficiency (deviation within 8%), while perturbation of the dipole source position within 150×150 nm$^2$ leads to the unstable performance of the designed Si$_3$N$_4$-hBN coupler.

DIMENSIONALITY REDUCTION ANALYSIS

A commonly used strategy with an adjoint TO framework is executing the same optimization algorithm with different initial conditions many times and picking the most efficient solution within the obtained set. However, this approach leads to an apparent tradeoff between time/resource and target performance. Such a dilemma arises since randomly selected initial conditions might lead to low-efficiency designs at the end of the time-consuming optimization run. Thus, there is a critical need to build a comprehensive understanding of what essential geometrical features should be included initially to increase the success rate of the TO. To build such an intuition, we applied a dimensionality reduction technique to the optimized design datasets and analyzed a clustering of the high-efficiency TO designs based on their geometrical features. Specifically, we apply t-distributed stochastic neighbor embedding (t-SNE), which is well suited for embedding high-dimensional data for visualization in 2D/3D-dimensional spaces.[33] This method allows us representing each TO design by a two-dimensional point and form clusters of such points representing designs with similar geometrical features. Analyzing the correlation between such clusters and corresponding design efficiencies, it is possible to gain insight into the critical geometrical features of the coupler design that lead to increased performance.

We use coupler design patterns as high-dimensional data; the corresponding coupling efficiencies are used as labels. To make the t-SNE analysis more meaningful, all the TO designs are grouped into 4 main classes based on the design's coupling efficiency by the k-means clustering method. This method partitions all the classes into a reasonable number of clusters in which each label belongs to the cluster with the nearest mean (centroids).[34]

After assigning a class to each design, we employed the t-SNE algorithm and obtained dimensionality-reduced distributions for "on top" (Fig. 3a) and "embedded" (Fig. 3b) configurations of the coupler. We found that all of the designs group into 4 clusters based on their geometrical similarity. It is important that this clustering is a good correlation with assigned efficiency classes, which indicates a connection between the efficiency of the coupler and its shape/topology. The efficiency classes are denoted with different marker colors. The classes are assigned based on the efficiency of the coupler. A visual and simple statistical analysis is applied to t-SNE plots to develop design rules for future optimizations. We formulate the design rules as two essential procedures, (i) enforcing material distributions in the selected locations and (ii) providing intuitive initial material density distributions (Fig. 3c). The first constrain restricts the material distribution in certain areas to remain intact throughout the optimization. These regions include (a) the proximity of the SPS (the red square, Fig. 3c-d), and (b) the cross-section connecting the SPS and the waveguide (the red segment, Fig. 3c-d). The details of this analysis, along with the application of machine learning algorithms, are presented in Supporting Information, Section S4.

Based on this analysis, we conclude that two regions of the initial material distributions play the critical role: (1) a Bragg reflector-like structure behind the emitter position (blue highlight, Fig. 3c), and (2) Si$_3$N$_4$ distribution within the area connecting the position of the emitter and the waveguide (green highlight, Fig. 3c). Figure 3c shows an example of the initial material distribution used for TO. The TO-related evolution of the coupling efficiency and the convergence of the Purcell factor yields 70% and 1.68 values, respectively (Fig. 3d). We found that the initial condition of Fig. 3c, chosen by the design rules, obtained from the t-SNE analysis, enables fast and stable convergence to highly efficient coupler designs (Fig. 3d). To prove this point, we conduct 25 TO runs for "on top" and "embedded" configurations. Figures 3e and 3f compare the coupling efficiencies of designs obtained from TO with either t-SNE-based or random sampling-based initial conditions. There is no surprise that the initial conditions are indeed similar to the structures already proposed in the literature for integrated SPSs with Bragg reflectors and tapered waveguide cross-sections.[35,36] This provides an excellent route to speed-up and ensure robust convergence through a physics-driven initial geometry fed into the TO framework. We note that such an initial "educated guess" could be done classically without the t-SNE and then perfected via TO.

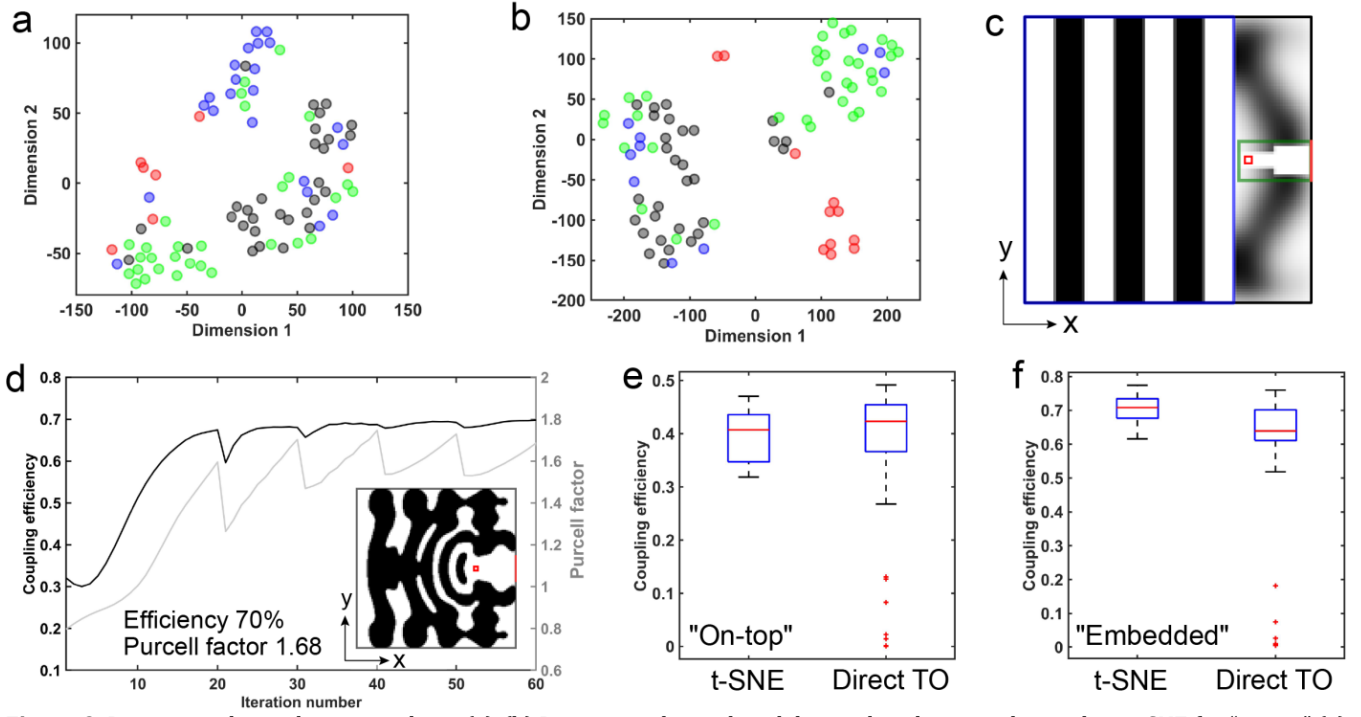

**Figure 3.** Dimensionality reduction analysis. (a)-(b) Dimensionality reduced design distributions obtained via t-SNE for "on top" (a) and "embedded" configurations. For "on-top" configuration, the classes are defined as(a): class 1: efficiency less than 10% (red markers), class 2: efficiency between 10% and 35%(blue), class 3: efficiency between 35% and 40%(green) and class 4: efficiency higher than 40%(black). For "embedded" configuration these classes are defined as (b): class 1: efficiency less than 10% (red markers), class 2: efficiency between 10% and 55% (blue), class 3: efficiency between 55% and 65% (green) and class 4: efficiency higher than 65% (black). (c) Initial material distribution obtained via t-SNE analysis. Blue highlight: a Bragg reflector-like structure behind the emitter position; green highlight: $Si_3N_4$ distribution within the area connecting position of the emitter and the waveguide; red square: position of the dipole source, surrounded by a constant material density of 1. (d) Convergence plots of the TO with the initial condition shown in (c). Black curve: FOM, grey curve: Purcell factor. Insets show final design of the $Si_3N_4$-hBN coupler. (e)-(f) Convergence plots of the TO with initial material distributions chosen based on design rule derived from t-SNE analysis("t-SNE") and random initial condition ("Direct TO") for "on-top" (e) and "embedded" (f) configurations. Box plot shows the median (red line), interquartile range (box) and outliers (red markers).

We found that in the case of the "on top" coupler, the t-SNE case has a much narrower efficiency distribution, while the maximum efficiency is slightly less (47%) than that in the case of direct TO (49%) with random sampling. The median of the t-SNE case distribution is 41% vs. the direct TO-produced distribution with a 42% median value. In the "embedded" case, the t-SNE-based initial condition gives a much better coupling efficiency distribution (71% median and 78% maximum efficiencies) vs. the direct TO ensures only 64% median and 76% maximum efficiencies. Hence, t-SNE based analysis indeed leads to a better convergence of the TO algorithm.

CONCLUSION

The main high-efficiency QPIC's components, such as SPE couplers, linear and nonlinear photonic components for quantum state manipulation, and out-couplers, are the focus of intensive research due to their importance in realization of novel quantum information processing applications. Along with efficiency, compatibility with the fabrication precision of state-of-the-art fabrication technology is one of the main constraints for the design and development process. In this work, we developed an optimization framework to design efficient structures to couple the emission from a point defect in 2-D materials, specifically, hBN, acting as a single photon emitter SPS to a photonic waveguide using adjoint topology optimization. The proposed TO method provides significantly enhanced coupling efficiency as well as drastically improved Purcell factors. We optimized the couplers for two different materials configurations and created a library of designs to gain deeper insights into the optimal device's performance.

We also demonstrated that physics-informed, intuitive initial conditions ensure rapid and robust optimization and produce high-performance devices that could pave the way to practical realizations of room temperature SPSs. Our approach enables fast and robust convergence with a physics-driven "educated guess" – an initial material distribution that could be obtained without the t-SNE, but perfected with the TO. Overall, the highest performing device achieves a coupling efficiency of 78% with nearly two times enhanced Purcell factor. As the following steps, fabrication imperfections and color center location uncertainties should be studied, and experimental realization of devices should be explored. The developed design approach can be extended to coupling schemes utilizing different SPSs as well as other inverse design problems in photonics.

ACKNOWLEDGMENTS

This work is supported by the US Department of Energy (DOE), Office of Science through the Quantum Science Center (QSC), a National Quantum Information Science Research Center (developing ML algorithms), DARPA/DSO Extreme Optics and Imaging (EXTREME) Program (HR00111720032) (OY, AVK, ZAK, TO techniques) and National Science Foundation award 2029553-ECCS.

# SUPPLEMENTARY INFORMATION

## S1 Setup for the Forward and Adjoint Simulations

Direct solver for full-wave numerical simulations in this study employs the finite-difference time-domain (FDTD) approximation (a commercial software suite, Lumerical FDTD is used). Configuration layouts for forward and adjoint simulations are shown in Figure S1. Both configurations use monochromatic sources with an excitation wavelength of 638 nm, matching the hBN zero-phonon emission line.

The forward simulation uses a dipole source to emulate the single-photon emitter. The lateral area of the coupler optimization domain is 2×2 μm$^2$ with a height of 200 nm. The coupler domain is connected to a waveguide with a matching height of 200 nm and a width of 500 nm. $E$- and $H$-fields at the waveguide are recorded with the monitor placed across the waveguide. At the adjoint simulation run, the dipole source is removed, and a perfectly matched waveguide mode source is placed at the monitor location from the forward simulation. The source excites only the fundamental mode of the waveguide. The amplitude of the mode excited by the source is determined from the coupling to the fundamental mode in the forward simulation. $E$- and $H$-fields in the coupler domain are recorded in both simulations and passed to the optimization engine for further calculations. The substrate and core materials are $SiO_2$ and $Si_3N_4$. Refractive index values are taken as 1.48, 2.1, 2 for $SiO_2$, $Si_3N_4$, and hBN, respectively.

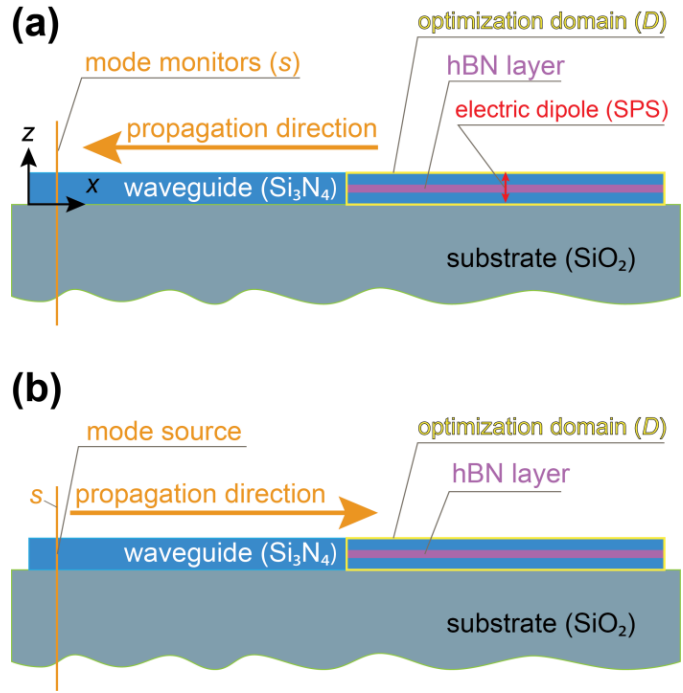

**Fig. S 1.** Configuration layouts for forward and adjoint simulations. (a) Forward simulation layout. The source for the forward simulation is an electric dipole (vertical red line with arrowheads), radiating energy into optimization domain $D$ (purple box). $E$- and $H$-fields coupled to the waveguide are recorded by the mode monitors placed across the waveguide (orange line). Power coupling to the desired mode is also calculated at the mode monitor. Forward fields used for the optimization are also recorded in the optimization domain. (b) Adjoint simulation layout. The source for the adjoint simulation is a waveguide mode with the amplitude retrieved from forward simulation. Field propagation is reversed with respect to the forward simulation setup. Adjoint $E$-fields for optimization are recorded in $D$.

## S2 Adjoint-Based Optimization for Single-Photon Emitter Couplers

Adjoint topology optimization is a powerful design technique leveraging the inherent symmetry in a general Maxwell operator for a reciprocal electromagnetic system.[1] This section presents the mathematical framework for the adjoint topology optimization used in our inverse design procedure for a high-efficiency single-photon emitter coupler. More general derivation for the adjoint method is available in the literature.[2,3] The adjoint method uses only two simulations (forward and adjoint) per optimization iteration to calculate gradients of the dielectric function across the desired optimization area. The main goal of the optimization is to maximize the coupling efficiency between the single-photon emission energy with the fundamental mode of the waveguide. Hence, the coupling efficiency is employed as the optimization figure of merit (FoM).

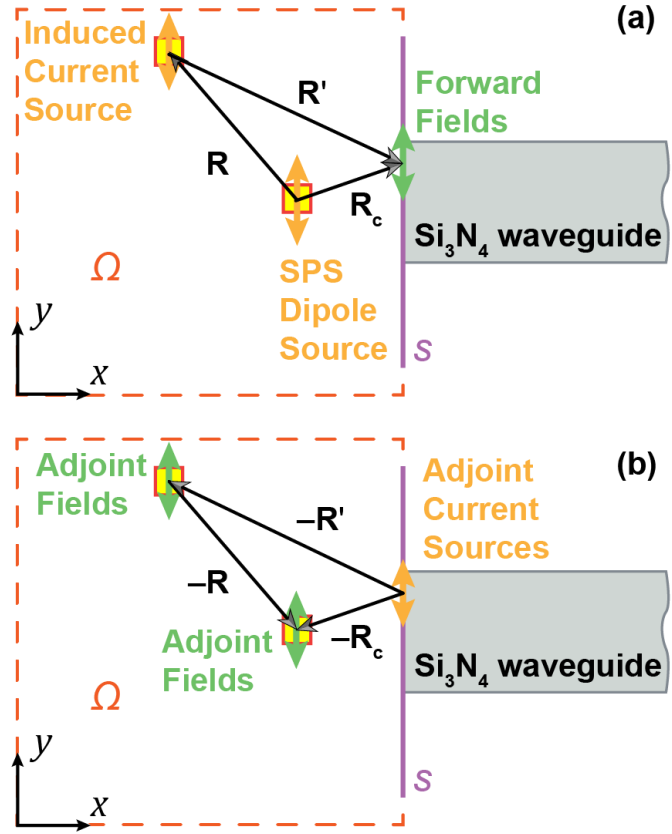

**Fig. S 2.** The forward (a) and adjoint (b) simulations schematic: these two steps are sufficient at each iteration. (a) The forward step: a single photon source (shown in orange) is excited inside the coupler. The induced dipoles are also shown in orange. The forward fields are obtained at cross-section $\Omega$. FoM is calculated at the waveguide cross-section $s$. (b) The adjoint step: the adjoint source is excited at the waveguide cross-section $s$. The induced adjoint fields are shown in green.

**Material density matrix.** Reactive etching fabrication method adopted for the device restricts its geometry to multilayer shapes with translational symmetry in the vertical $z$-direction. Hence, we employ a density-based material parametrization following the onset discussed in[4,5], and utilize a 2D matrix of normalized (variation-limited) design parameters,

$$\rho = \begin{bmatrix} \rho_{11} & \cdots & \rho_{i_{max}1} \\ \vdots & \ddots & \vdots \\ \rho_{1j_{max}} & \cdots & \rho_{i_{max}j_{max}} \end{bmatrix} \tag{S2}$$

Matrix $\rho$ is used to map the normalized distribution over cross-section $\Omega$, shown by the orange dashed box in Fig. 1(a) into the discretized permittivity distribution inside the entire coupler. Each term $\rho_{ij} = \rho(\mathbf{r}_{ij})$ of the material density matrix has a limited variation from 0 (air) to 1 (material), with $\mathbf{r} \in \Omega$ being a 2D position vector in $\Omega$.[42,4–7]

**Coupling Efficiency (FoM).** The normalized coupling efficiency is defined as the ratio of the squared electromagnetic energy flux converted into the lowest-order power-normalized TEM mode to the Poynting flux passing through the surface of a small cubic volume encapsulating the SPS,

$$C = \frac{|P|^2}{P_{SPS} P_0} \tag{S3}$$

where

$$P = \tfrac{1}{4} \int_s \left( \mathbf{E} \times \mathbf{H}_0^* + \mathbf{E}_0^* \times \mathbf{H} \right) \cdot \hat{\mathbf{z}} \, ds \tag{S4}$$

and

$$P_{\text{SPS}} = \tfrac{1}{2} \oint_{S_0} \Re\left[\mathbf{E}_{\text{d}} \times \mathbf{H}_{\text{d}}^*\right] \cdot d\mathbf{S}_0 \tag{S5}$$

$P$ and $P_{\text{SPS}}$ are the power converted into the fundamental waveguide mode and power injected by the SPS into the simulation domain, respectively. $\mathbf{E}_{\text{d}}$ and $\mathbf{H}_{\text{d}}^*$ are the fields emitted by the dipole into the simulation domain, calculated at the surface of a small cubic shell ($S_0$) encapsulating the dipole. $\mathbf{E}, \mathbf{H}$ fields generated by the SPS, and $\mathbf{E}_0, \mathbf{H}_0$ are the fields of the fundamental TEM mode pattern at the monitor cross-section area *s*. The overall objective of the TO cycle is obtaining the material density distribution (2D matrix $\boldsymbol{\rho}$) in $\Omega$, mapped into the permittivity distribution in optimization domain *D,* maximizing the coupling efficiency,

$$\max_{\boldsymbol{\rho}} C(\mathbf{E}) \tag{S6}$$

**Filtering out the fundamental mode.** We separate the power converted into the fundamental waveguide mode by presenting the fields ($\mathbf{E}, \mathbf{H}$) calculated at each iteration step, as a combination of the scaled fields of the fundamental mode ($a\mathbf{E}_{\text{f}}$, $a\mathbf{H}_{\text{f}}$), with $a$ being a scaling factor, and some other modes $\mathbf{E} = a\mathbf{E}_0 + \mathbf{E}_\Sigma$ and $\mathbf{H} = a\mathbf{H}_0 + \mathbf{H}_\Sigma$,

$$\begin{aligned} P &= \tfrac{1}{4} \int_s \left[\mathbf{E} \times \mathbf{H}_0^* + \mathbf{E}_0^* \times \mathbf{H}\right] \cdot \hat{\mathbf{x}} ds \\ &= a\tfrac{1}{2} \int_s \Re\left[\mathbf{E}_0 \times \mathbf{H}_0^*\right] \cdot \hat{\mathbf{x}} ds + \tfrac{1}{4} \underbrace{\int_{s_c} \left[\mathbf{E}_\Sigma \times \mathbf{H}_0^* + \mathbf{E}_0^* \times \mathbf{H}_\Sigma\right] \cdot \hat{\mathbf{x}} ds}_{\text{orthonormality}}. \end{aligned} \tag{S7}$$

The second integral in (S7) vanishes due to the classical orthonormality condition between the fundamental and other waveguide modes.[8] Hence, the initial integration of the calculated fields with ($\mathbf{E}, \mathbf{H}$) the normalized fundamental mode patterns ($\mathbf{E}_{\text{f}}, \mathbf{H}_{\text{f}}$) leaves only the power of the fundamental mode, where the fundamental mode patterns are normalized to give,

$$P_0 = \frac{1}{2} \int_s \Re\left[\mathbf{E}_0 \times \mathbf{H}_0^*\right] \cdot \hat{\mathbf{x}} ds = 1\,[\text{W}]. \tag{S8}$$

**Induced field.** It is assumed that a small perturbation of permittivity $\delta\varepsilon_{ijk}$ excited with a local *E*-field generates a perturbation of the associated current density ($\delta\mathbf{J}_{ijk}$). For simplicity, the current density perturbation is taken directly proportional to the perturbation of voxel permittivity $\delta\varepsilon_{ijk}$,

$$\delta\mathbf{J}_{ijk} = -\iota\omega\varepsilon_0 \left[\delta\varepsilon_{ijk}\mathbf{E}_{ijk}\right] \tag{S9}$$

Then, a linear operator $\mathbf{G}_{ijk}$ can be used, connecting the perturbation of inducing source[9] $\delta\mathbf{J}_{ijk}$ with the perturbed induced field data $\delta_{ijk}\mathbf{E}$,

$$\begin{aligned} \delta_{ijk}\mathbf{E} &= \mathbf{G}_{ijk}\delta\mathbf{J}_{ijk}, \\ \mathbf{R}' &= \mathbf{R}_c - \mathbf{R}_{ijk} \end{aligned} \tag{S10}$$

here $\mathbf{G}_{ijk}$ is a general linear operator connecting the induced *E*-fields with inducing sources $\mathbf{J}_{ijk}$, respectively. The form of operator $\mathbf{G}_{ijk}$ may vary depending on the actual solver used for calculating the field and could take different forms for the FEM[3,4], FDFD[10], and other numerical techniques[11].

**Efficiency gradient.** As its name implies, a gradient-based optimization (S6), is built on computing the gradient of *C*. First, we connect the normalized field patterns $\omega\mu_0\mathbf{H}_0^* = \beta\hat{\mathbf{x}} \times \mathbf{E}_0^*$ (where $\beta = k_0 k_z$ is a propagation constant). Then, we assume that a small perturbation of permittivity ($\delta \varepsilon_{ijk}$) produces a small change to the E-field ($\delta_{ijk}\mathbf{E}$) at the waveguide cross-section $S_c$, $\mathbf{E} = a\mathbf{E}_0 + \delta_{ijk}\mathbf{E}$, and after neglecting the quadratic term, we get,

$$\delta_{ijk}C \approx \frac{2\Re\left(\int_s \delta_{ijk}\mathbf{E} \cdot A\mathbf{E}_0^* \mathrm{d}s\right)}{P_{\mathrm{SPS}}}, \quad A = a\eta_0 k_z \tag{S11}$$

with $\eta_0 = \sqrt{\varepsilon_0/\mu_0}$.

**The forward step: the scaling factor.** Filtering procedure (S7) returns scaling coefficient $a = P/P_0$ at each forward step of the optimization cycle. Substituting (S10) into (S7) gives a perturbation of power generated by a perturbed voxel (with induced current density $\mathbf{J}_{ijk}$) at the monitor cross-section $s$. Since for any $\mathbf{G} = \begin{bmatrix} \mathbf{g}_x & \mathbf{g}_y & \mathbf{g}_z \end{bmatrix}^{\mathrm{T}}$, $\mathbf{G} \in \mathbb{R}^{3\times 3}$, $\mathbf{v},\mathbf{w},\mathbf{g}_i \in \mathbb{R}^3$, $i = \{x,y,z\}$, the following simplifying identity holds, $\left[\mathbf{G} \otimes \mathbf{v}\right] \cdot \mathbf{w} = \mathbf{v} \cdot \left[\mathbf{G}^{\mathrm{T}} \otimes \mathbf{w}\right]$. Then,

$$\frac{\delta_{ijk}C}{\delta \varepsilon_{ijk}} \approx \frac{-2\Re\ \mathbf{E}_{ijk} \cdot \mathbf{E}_{ijk}^{adjo\,\mathrm{int}}}{P_{\mathrm{SPS}}}, \quad \mathbf{E}_{ijk}^{adjo\,\mathrm{int}} = \int_s \mathbf{G}_{ijk}^{\mathrm{T}} \otimes A\mathbf{J}_0^*\ \mathrm{d}s, \tag{S12}$$

where $\mathbf{J}_0^* = \iota\omega\varepsilon_0 \mathbf{E}_0^*$ is the adjoint current density generated by the adjoint E-field strength by the fundamental waveguide mode with a scaled field pattern $\mathbf{E}_0^*$. The integral in (S12) uses reciprocity to compute the adjoint field at the perturbed location ($\mathbf{E}_{ijk}^{adjo\,\mathrm{int}}$).

The actual way of reconstructing the adjoint field in the device domain $D$ from the field pattern depends on the selected solver.[12] In our case, a commercial FDTD solver has been used to obtain all the forward and adjoint fields. Equation (S12) gives the partial derivatives of $C$ with a clear path to adjoint topology optimization (see, Section S3, page 12).

**Purcell Factor.** Since the overall emission is largely modified by the dielectric composition around the emitter, the Purcell factor is also affected by the optimization process. Purcell factor here is defined as the ratio of light flux going through a small cubic volume around the source between the optimized and the homogenous material distributions.

$$\mathrm{F}_P = \frac{P_c}{P_\mathrm{h}} = \frac{\frac{1}{2}\oint_{s_0} \Re\ {}^c\mathbf{E}_\mathrm{d} \times {}^c\mathbf{H}_\mathrm{d}^*\ \cdot \mathrm{d}\mathbf{s}_0}{\frac{1}{2}\oint_{s_0} \Re\ {}^h\mathbf{E}_\mathrm{d} \times {}^h\mathbf{H}_\mathrm{d}^*\ \cdot \mathrm{d}\mathbf{s}_0} \tag{S13}$$

$P_\mathrm{h}$ and $P_\mathrm{c}$ are the time-averaged energy density fluxes going through a small box encapsulating the source inside optimization domain $D$, filled with either homogenous $Si_3N_4$ ($P_\mathrm{h}$) or optimized binary $Si_3N_4$-Vacuum composite ($P_\mathrm{c}$).

## S3 Iterative Optimization Procedure

The adjoint topology optimization is an iterative inverse design process typically lasting several hundred iterations for a design to converge. The number of iterations is strictly controlled by the gradient scaler ($\alpha$), binarization push coefficient, filter radius, and initial condition. One can gain intuition into the proper selection of these parameters after a few initial runs. The optimization process is summarized in Fig. S 3. In the rest of this section, we follow the flowchart and lay out the details of each step.

**Selecting the Initial Material Density Distribution.** Initial material distribution is one of the most impactful factors in topology optimization since the gradient descent nature of the procedure finds the local minima which is dictated by the starting point in the optimization hyperplane.[13] The initial distribution can either be set randomly or based on intuitive topologies for the specific design. In addition, the intuition behind the high-performance topologies can be deduced by inspecting the material distribution of the high-efficiency devices, which in return can inspire initial material distribution yielding even better performance characteristics. For the first part of this study, we randomly sampled the optimization plane for initial material density distribution. Later, we used intuitive starting points based on the high-performance topologies. Details of this procedure is discussed in Section S4.

**Filtering and Binary Push.** During the topology optimization, the dielectric constant at all spatial voxels in the device can range in continuous fashion between selected material platforms. The continuous distribution of the refractive index is realized by the material density matrix. The optimization procedure modifies the density matrix, which is later used to calculate the exact permittivity at each discrete voxel.

$$\begin{aligned}\varepsilon_{\text{air}} &< \varepsilon_{ij} < \varepsilon_{mat} \\ 0 &< \rho_{ij} < 1 \\ \varepsilon_{ij} &= \rho_{ij}\varepsilon_{mat} + (1-\rho_{ij})\varepsilon_{\text{air}}\end{aligned} \quad (S14)$$

where $\rho_{ij}$ is the material density distribution that varies from 0 (air) to 1 (material); $\varepsilon_{\text{air}} = 1$ and $\varepsilon_{\text{mat}}$ are the dielectric constants of air and the material domain (Si$_3$N$_4$ and hBN), respectively, $\varepsilon_{\text{mat}} \in \{\varepsilon_{\text{Si}_3\text{N}_4} = 2.1^2, \varepsilon_{\text{hBN}} = 4\}$.

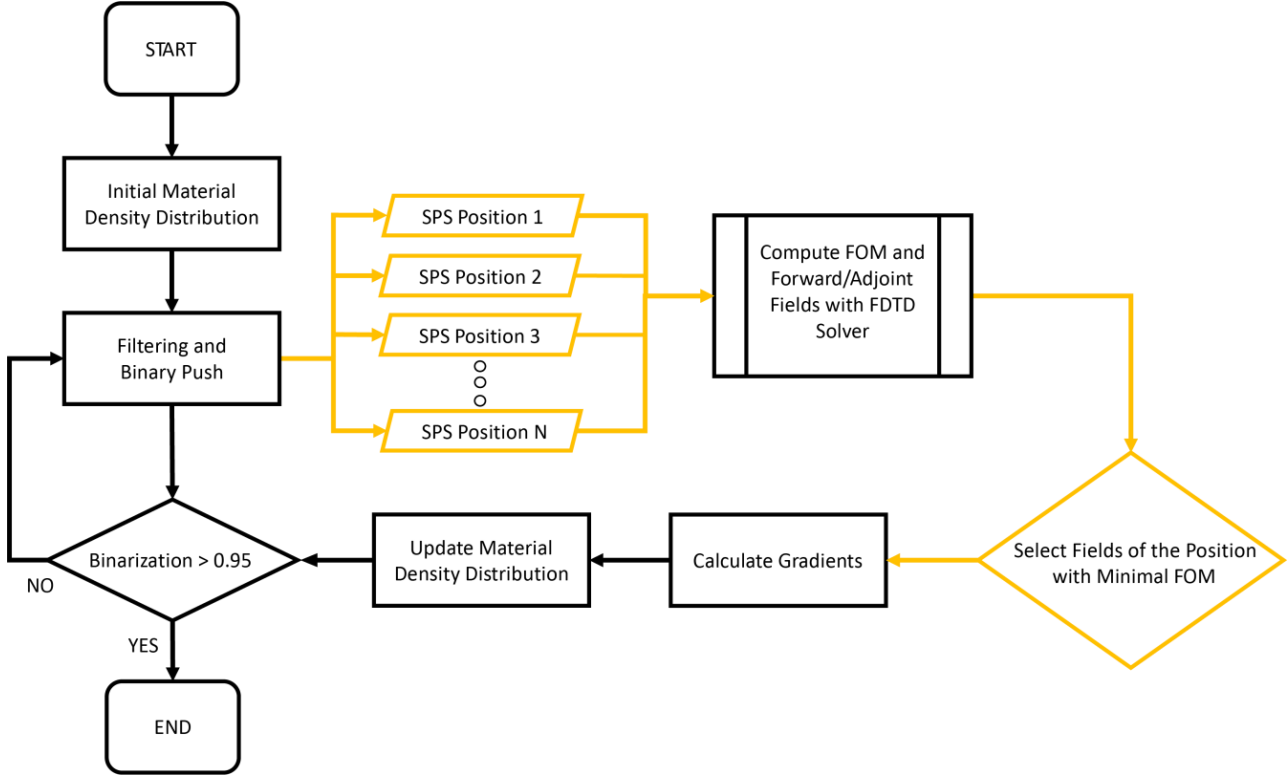

**Fig. S 3.** Flowchart of the optimization procedure.

Feature sizes in any photonic device are constrained by fabrication capabilities. To conform with state-of-the-art fabrication methods such as electron beam lithography, features with dimensions below a certain threshold need to be removed. These features include tiny protrusions from the central bodies of small islands of material/hole. To this end, we applied a circular spatial blurring filter to the material distribution every few tens of iterations,

$$ij \in \mathbb{N}_e^{2\times 2}, \; if \; \|\mathbf{r}_{ij} - \mathbf{r}_e\| \leq r_{filter}$$
$$\rho_e = \frac{\sum\limits_{ij\in\mathbb{N}_e^{2\times 2}} \rho_{ij}\left[r_{filter} - \dfrac{\|\mathbf{r}_{ij}-\mathbf{r}_e\|}{r_{filter}}\right]}{\sum\limits_{ij\in\mathbb{N}_e^{2\times 2}}\left[r_{filter} - \dfrac{\|\mathbf{r}_{ij}-\mathbf{r}_e\|}{r_{filter}}\right]} \quad (S15)$$

where $r_{filter}$ is the filter radius, $\mathbb{N}_e^{2\times 2}$ is the set containing the grid points within the filter radius with respect to a given filtered point $\rho_e$. The filter radius is selected to be 60 nm for this study. The application of the filter reduces the FoM immediately since the optimized density matrix is blurred. However, this procedure is applied multiple times until a converged design minimally compromised by the filtering procedure emerges. In rare cases, the filtered out material distribution can even increase the performance. In addition, we apply hard constraints over the density matrix in a square area enclosing the diploe source. Since uncertainties in the SPS location are expected in the experiment, the close vicinity of the source (a 60×60 nm$^2$) remains intact throughout the entire optimization process.

To converge to a fully-binarized refractive index distribution, we apply *the binary push* to the material density matrix in the form of a Heaviside filter,[14,15]

$$\varepsilon_{ij} = \frac{\tanh(\gamma\sigma) + \tanh[\gamma(\rho_{ij} - \sigma)]}{\tanh(\gamma\sigma) + \tanh[\gamma(1 - \sigma)]} \tag{S16}$$

where $\gamma$ and $\sigma$ are the coefficients determining the strength and direction of the binary push. This step ensures that the design is converged within reasonable amounts of iterations. The optimization stops when the overall binarization of the spatial voxels is above 95%. After this point, un-binarized voxels are forced to take binary values by thresholding.

**Changing the SPS position (optional):** Localization of SPS in host medium is a challenging task which is often diffraction-limited by optical detection systems.[16] Therefore, the optimization scheme has a robustness implementation in which the SPS (forward simulation source) is randomly placed in tangential directions with pre-selected limits in length multiple times at each iteration. The number of the forward simulation at each step is dictated by amount point which is required to be scanned. All the forward fields along with their respective FoMs are recorded separately for gradient calculation in the next steps. We note that the step explained here is optional and only necessary for enabling the design robustness.

**Computing the FoM, Forward and Adjoint Fields with the FDTD solver:** The overall optimization procedure is done using the Matlab scripting language as an application programming interface coupled with a commercial finite-difference time-domain solver (Lumerical FDTD). Fields related to forward/adjoint simulations and FoM are recorded with field monitors at the waveguide cross-section *s* and optimization domain *D*. The FoM is computed from equation (S4).

**Selecting Fields for the SPS position with the Minimal FoM (optional):** To achieve robustness against SPS displacement, TO procedure must ensure that the worst location in terms of FoM within the uncertainty limits yields reasonably well-performing devices. Therefore, maximin decision rule is applied as the forward field selection criterion where the fields of the location with the minimal FoM are selected for gradient calculation.

**Calculating the Gradients:** Recorded fields in the optimization area from forward/adjoint simulations are 3D data structures. However, we are only interested in multilayer structures with translation symmetry in vertical direction. We average the fields along the vertical *z*-direction to account for contribution from each layer in *D*. After collapsing the 3D *E*-fields into 2D matrix, the gradient is calculated through,

$$\nabla_\rho C_{ij} = \sum_{ij} \partial_{\varepsilon_{ij}} C_{ij} \nabla_\rho \varepsilon_{ij} \tag{S16}$$

The first multiplicand on the left-hand side is calculated with (S12) while the second one is obtained by applying the chain rule to (S14)-(S16).

**Updating the Material Density Distribution:** At every iteration, calculated fields inside the 3-D optimization area are collapsed to the 2-D matrix to match matrix $\boldsymbol{\rho}$ by averaging the fields in FDTD mesh grid points in the vertical *z*-direction. Therefore, material composition is identical in this direction inside the optimization area. Then, these fields are used to calculate gradients ($\nabla_\rho C_n$) which are scaled by the selected amount and added to the index density matrix ($\boldsymbol{\rho}_n$).

$$\boldsymbol{\rho}_{n+1} = \boldsymbol{\rho}_n + \alpha \nabla_\rho C_n \tag{S17}$$

**Stop Condition (Binarization > 95%):** Once ratio of the number of voxels with dielectric constants equal to Air/$Si_3N_4$ to all voxels reaches a desired threshold (95% for this study), the material density matrix is completely binarized by thresholding all the voxels and forcing them to assume a binary value. This condition greatly helps with time reduction since reaching a fully-binarized distribution without extreme binarization coefficients can take significant number of iterations where the same structure with virtually no changes is calculated repeatably.

## S4   Dimensionality Reduction Analysis and Design Rule Derivation

Adjoint TO is gradient-descent based optimization method which converges to the local minima points in the parameter hyperplane. Therefore, initial material density distribution is critical to fully-optimized design performance. For the particular case of this study, an intuitive structure for high coupling efficiency is shown in Fig. S4. A tapered waveguide cross-section ensuring the impedance match between the SPS and the fundamental mode of the waveguide is placed in the area connecting the two and a Bragg-like reflector reflecting and channeling energy into the waveguide is placed at the back of the optimization area. Intuition of similar nature can be obtained through deep exploration of the topologically optimized couplers, which can be used to access even higher-performing devices in shorter design times.

Using the inverse design method outlined above, we produce 2 datasets containing 86 fully optimized devices for two different configurations of the HBN flake. The optimized designs are handled as 201×201-pixel images, whereas their figure of merit (coupling efficiency) becomes their labels. Each device (image) has a different performance (label), so that grouping them directly would not highlight their similarities. Therefore, we quantized the vector of performance labels by *k*-means clustering.[17] This method partitions arbitrary amounts of observations into the desired number of clusters where each observation belongs to a cluster with the nearest mean. We found out that having 4 centroids for both datasets are sufficient. We followed different approaches for the different datasets. For the "embedded" configuration, we directly applied t-SNE to the dataset with 8 perplexities, a learning rate of 400, and 1000 iterations. For the "on top" configuration, we started with a PCA (Principal Component Analysis) initialization. We took projections of the images on 5-element vectors. Later, we applied t-SNE with 10 perplexities, a learning rate of 200 within 1000 iterations.

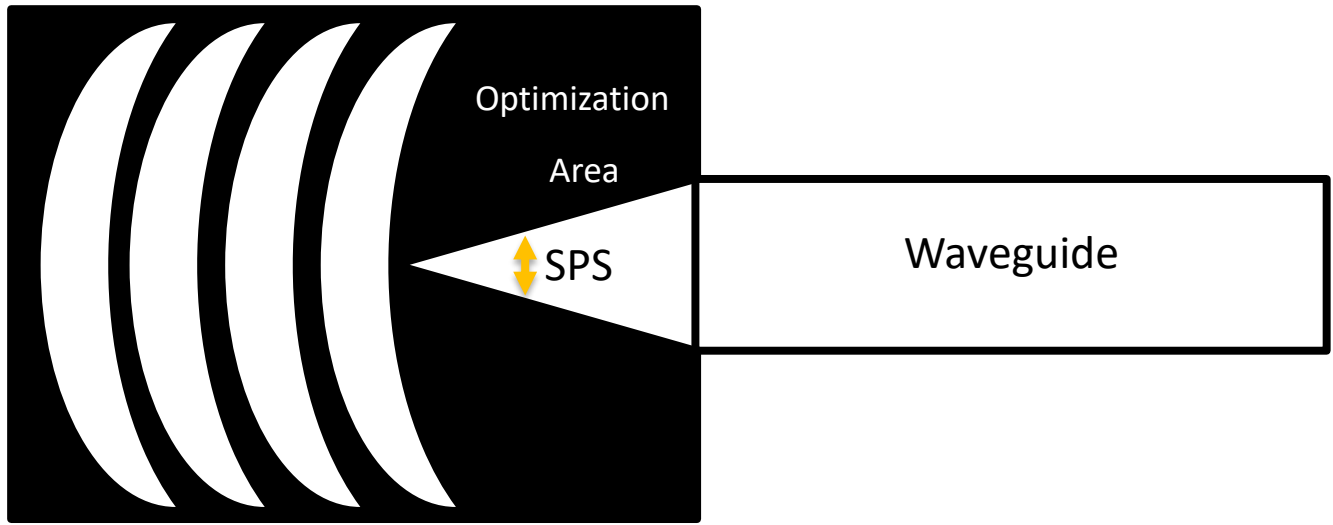

**Fig. S 4.** Top view artistic rendition of an intuitive human design for the SPS-to-waveguide coupler. Black and white indicate the areas filled with air and $Si_3N_4$, respectively.

After creating meaningful t-SNE maps, we performed a manual analysis. In this procedure, we essentially performed simple matrix operations between the images, such as averaging a cluster in a t-SNE plot or locating mutual material/hole features between the samples by intersection. After a comprehensive analysis step, we arrived at the design rules presented in the paper. We created smaller datasets with the derived design rules where all the devices are high performing compared to original datasets. The design rules are applied in the form of (i) enforced material distributions in the selected locations and (ii) intuitive initial material density distributions, shown in Fig. 3 of the main text. The enforced material distribution in certain areas does not change throughout the optimization. These regions are mainly (a) the proximity of the SPS (see the red square in Fig. 3cd of the main text), and (b) the cross-section connecting the SPS and the waveguide (the red segments in Figs. 3cd of the main text).